\documentstyle[aps,prl,epsf,color]{revtex}

\begin{document}

\unitlength1.0cm
\draft

\title{Hollow density channels and transport in a laser irradiated 
plasma slab}

\author{H. Ruhl}
\address{Max-Born-Institut, Max-Born-Stra\ss{}e 2a, 12489 Berlin, Germany} 

\date{\today}

\maketitle

\begin{abstract}
A three dimensional Particle-In-Cell simulation describing the
interaction of an intense laser beam with a plasma slab is 
presented. It is observed that the laser generated electron 
current decays into magnetically isolated filaments. The 
filaments grow in scale and magnitude by a merging process
in the course of which the field topology changes. The opposite 
process also takes place occasionally. The laser driven charge 
and energy flows and the reconnecting magnetic field mutually 
interact. At the end of the merging process flows and fields
are confined close to the laser irradiated surface of the 
plasma slab. Both decay rapidly in the bulk plasma. Due to
the magnetic pressure in the filaments hollow density channels 
in the electron and ion densities are formed. The simulation
reveals that charge flows in these channels can exceed the 
Alfven current.
\end{abstract}


Key issues of many applications concerning the interaction between
lasers and matter at high intensities are laser absorption and 
charge and energy transport through a plasma. A typical application 
of intense laser-matter interaction where this is of relevance is 
Fast Ignition (FI) in Inertial Confinement Fusion (ICF) 
\cite{TabakPHP94,HainPRL01,RothPRL01}. In the present paper we study
issues of transport in a thick plasma slab with the help of 
Particle-In-Cell (PIC) simulations over a few hundred femtoseconds
simulation time. Binary collisions are included. Details of the 
collisional model will be presented elsewhere. The density of the 
plasma slab is about $20$ times over-critical.

A reasonable definition of thickness has some ambiguity. We call 
a slab thick when its depth exceeds many times the skin lengths 
$l_{\text{s}}=c/\omega_{\text{p}}$ where $\omega_{\text{p}}$ is 
the plasma frequency. In addition, the thickness of the plasma
slab must be larger than the penetration depth of the charge
flow. The plasma slab investigated in this paper has a thickness 
which is large enough to exceed the latter sufficiently.
Two dimensional (2D) transport simulations in the plane perpendicular 
to the laser direction have recently been reported \cite{MTVPRL00}. 
In these simulations the slab thickness vanishes, the laser is 
neglected and fast particles have been injected by hand. In addition, 
2D geometry severely limits the available degrees of freedom for 
current transport and magnetic field evolution. 

In the present paper we investigate properties of laser-generated 
charge flows in a thick plasma slab in 3D where the fast particles 
are generated by laser irradiation. We show that magnetic filament merging 
\cite{Rosenbluth} is an important process to form large filaments 
and magnetic field strengths. In 3D the process represents a simple 
form of magnetic reconnection in the sense that the field topology 
changes in the course of the latter. Laser generated charge flows 
found in the simulation exceed the Alfven limit \cite{AlfvenPR39,LawsonJEC58}.
Furthermore, it is observed that the magnetic fields dominate the
properties of current transport. Both, electric and magnetic fields
are localized close to the front surface of the plasma slab as is 
the laser generated charge flow. 

The slab has a thickness of $6.0 \, \mu \text{m}$ and a width of $4.0 
\, \mu\text{m} \times 4.0 \, \mu \text{m}$. The box size is $4.0 \, 
\mu\text{m} \times 4.0 \, \mu\text{m} \times 8.0 \, \mu \text{m}$.
All fields depend on $x$, $y$, and $z$. The numerical grid has $152 \times 
152 \times 800$ cells. Lateral directions are periodic. Electrons and 
ions are presented by $1.44 \cdot 10^8$ quasi-particles. The initial 
electron and ion temperatures are $10.0 \, \text{keV}$ and $1.0 \, 
\text{keV}$ respectively. The laser beam propagates in $z$ and 
is linearly polarized along $x$. After a rise time of three optical 
cycles the laser intensity is kept constant. The incident laser beam
has a Gaussian envelope laterally with a width of $2.0 \mu \text{m}$ at 
full-width-half-maximum. The irradiance is $I\lambda^2=5.0 \cdot 10^{19} 
\text{Wcm}^{-2}\mu \text{m}^2$. The slab has a marginal initial 
deformation to enhance absorption \cite{RuhlPRL99}. The deformation 
is parameterized by $z(x,y)=\delta \cdot \exp \left( -(x-x_{\text{0}})^2/
r^2+(y-y_{\text{0}})^2/r^2 \right)$ where $\delta=0.4 \mu\text{m}$, 
$x_{\text{0}}=y_{\text{0}}=2.0 \, \mu \text{m}$ and $r=2.5 \, \mu\text{m}$. 
The center of the slab is located at $z=4.2 \, \mu \text{m}$. The 
background plasma consist of protons. The initial electron plasma density 
in the simulation is $n_{\text{e}}=3.33 \cdot 10^{22} \text{cm}^{-3}$. 
The laser radiation is turned off after $200 \, \text{fs}$. The
simulation itself is stopped after $400 \, \text{fs}$. The total
simulation is carried out on a parallel computer using $361$ compute
nodes and consumes $29000$ CPU hours in total. The coordinate system 
used in the simulations is right handed. Hence, the positive $z$-axis 
points out of the $xy$-plane shown in the figures. In what follows we 
will always use the positive $z$-axis for reference. 

Plots (a,b,c) of Fig. \ref{fig:bt} show the plane $z=2.0 \, \mu \text{m}$ 
of the cycle averaged magnetic field $|{\bf B}|$ defined in the figure 
caption. Time proceeds from plots (a,d) to (c,f). Filaments of different 
scale are observed. The white arrows indicate the direction of the magnetic 
field. It is seen that the magnetic field lines in the filaments are closed. 
As is seen from plots (a,b) the filaments attract each other. In case the 
attractive forces between them are strong enough magnetic field lines start 
to reconnect to form larger filaments with again closed field lines. The 
topology of the magnetic field changes. The inverse process
takes place occasionally. The largest filament scale in our simulation 
is shown in plot (c) of the figure. The peak magnetic field strength in 
this filament is about $45000 \, \text{T}$. Plots (d,e,f) show $\int dxdy 
\, |{\bf E}|$ and $\int dxdy \, |{\bf B}|$ obtained from the cycle averaged 
fields ${\bf E}$ and ${\bf B}$ in the simulation box. They illustrate 
the magnitude of ${\bf E}$ and ${\bf B}$ along $z$ in the bulk of the 
plasma. The plots show that close to the front surface and in the
bulk of the plasma slab the magnetic field ${\text{B}}$ is larger than 
the electric field $\text{E}$. At the rear of the slab $\text{E}$ is
larger than $\text{B}$. The electric field ${\bf E}$ at the rear of 
the slab points along the positive $z$-axis. At the front surface it 
points in the opposite direction. Furthermore, the relations $E_{\text{z}} 
\gg E_{\text{x,y}}$ and $B_{\text{z}} \ll B_{\text{x,y}}$ are obtained
in the simulation. Since the electric field is largest at the rear surface 
back surface acceleration of protons is most efficient in this simulation.

Magnetic and current filaments are very small initially. There have 
been efforts to describe this early stage with the help of the
Weibel theory \cite{CalifanoPRE97,SentokuPHP00}. When two filaments 
approach each other the magnetic fields of the latter tend to cancel 
in the overlapping region leading to attractive forces between the
filaments. This process is rapid in 3D and accompanied by a change 
of field topology. It cannot be recovered within the framework of
the Weibel theory.

Plots (a,b,c) of Fig. \ref{fig:jtevol} show the cycle averaged current 
density filaments $j_{\text{z}}$ at different times for slices at $z=2.0 \, 
\mu\text{m}$. Plots (d,e,f) of the same figure show the total current 
and return current in $z$-direction obtained from $I=\int dxdy \, 
j_{\text{z}}$. The peak current density in the filaments shown in plots 
(a,b,c) is $|j_{\text{z}}|=5.5 \cdot 10^{17} \, \text{A/m}^2$. The peak 
total current is about $I=7.0 \cdot 10^{5} \, \text{A}$. Total current 
and return current are of equal magnitude while there are large 
discrepancies between both in individual filaments. The total current 
is largest close to the front surface of the slab. It declines in the
bulk plasma. It has been split into its positive and negative components.
To estimate the Alfven current $I_{\text{A}}=4\pi\epsilon_0m_ec^3 \, 
\beta \gamma /e$ the quantities $\beta=v/c$ and $\gamma=1/\sqrt{1-\beta^2}$
are required. The simulation yields $\beta \approx 1.0$ and $\gamma 
\approx 2.0$ for the central filament in plot (c). This gives $I_{\text{A}}
\approx 34.0 \, \text{kA}$. However, the total current in the central 
filament is about $200 \, \text{kA}$ which exceeds the Alfven current 
almost $6$-fold. We note that the quantity $\beta$ has been calculated
from $\beta({\bf x},t)=\int d^3p \, v_{\text{z}} \, f({\bf x}, {\bf p},
t)/ c \, \int d^3p \, f({\bf x}, {\bf p},t)$ averaged over the 
cross-sectional area $A$ of the central filament in plot (c). This yields 
$\beta=\int dxdy \, \beta({\bf x},t)/A$. To make $\beta$ as large as 
possible only the fastest electrons are included in the calculation with
the requirement that the net current flowing in the filament is recovered
from the latter.

Figure \ref{fig:ne} shows the formation of hollow, honeycombed density 
channels in the electron density. The same are obtained in the proton 
density (not shown in the present paper). Plots (a,b,c) of the figure 
show slices at $z=2.0 \, \mu\text{m}$ for different times. Plot (d) 
gives a density slice at $x=2.0 \, \mu\text{m}$. In plot (d) a central
channel (blue color) surrounded by channels to the left and right of it 
is visible. These channels contain many hot electrons that exit the
rear surface of the plasma slab. The energy density inside the channels
is enhanced. The channels to the left and right of the central one coil 
around the latter. Hence, they are not fully visible in the plane at 
$x=2.0 \, \mu\text{m}$. Comparison of Figs. \ref{fig:bt} and Figs. 
\ref{fig:ne} reveals that the hollow electron density channels are 
generated by the pressure of the magnetic field.

In the context of current transport the singular current $I_{\text{s}}$ 
is of relevance. It is defined as the total current $I$ minus the guiding 
center current $I_{\text{gc}}$. The guiding center current comprises 
basically those electrons that gyrate around the magnetic field lines 
while the singular current contains mainly electrons that meander between 
field lines. The singular electrons are those that are capable of 
transporting large currents. The simulation yields approximate pressure 
balance between magnetic and thermal pressures on slowly varying time 
scales. Hence, we find

\begin{eqnarray}
\label{pressure_balance_integral}
\partial_i P_{ij} &\approx& \left( {\bf j} \times {\bf B} \right)_j 
\; , \qquad P_{ij}=\int d^3p \; p_i v_j f \; .
\end{eqnarray}

The quantity $P_{\text{ij}}$ in Eq. (\ref{pressure_balance_integral})
is the pressure tensor which is calculated directly in the simulations. 
Rewriting Eq. (\ref{pressure_balance_integral}) yields

\begin{eqnarray}
\label{pressure_balance_polar_coord}
j_z&\approx&\frac{1}{r} \, \partial_r \left( 
\frac{rP_{\perp}}{B_{\theta}} \right)+ \frac{P_{\perp}}{B^2_{\theta}} \,
\partial_{r} B_{\theta} \; .
\end{eqnarray}

Equation (\ref{pressure_balance_polar_coord}) is obtained from Eq. 
(\ref{pressure_balance_integral}) assuming cylindrical geometry
for a filament and rewriting $P_{\text{ij}}$ with the help of the 
pressure $P_{\perp}$ in radial direction normal to the $z$-axis. The 
magnetic field has been approximated by $B_{\theta}$ and the current 
density by $j_{\text{z}}$. Cylinder symmetry is approximately 
applicable to the situation found in the simulation. Hence, the 
calculation reveals the impact of geometry, pressure, and current 
density on the total current that can flow. Ambiguities inherent in 
the calculation of $I_{\text{A}}$ as discussed above are avoided. The 
singular current $I_{\text{s}}$ is obtained with the help of Eq. 
(\ref{pressure_balance_polar_coord})

\begin{eqnarray}
\label{singular_current}
I_s&=&I-I_{gc}=-4\pi \epsilon_0 c^2 \, \frac{P_{\perp}(0)}{j_z(0)} \; .
\end{eqnarray}

For a mono-energetic cold electron beam with electron density $n_{\text{0}}$
and velocity $v$ for which we have $P_{\perp}=m_{\text{e}}n_{\text{0}}v^2$ 
and $j_{\text{z}}=-en_{\text{0}}v$ the singular current $I_{\text{s}}$ in 
Eq. (\ref{singular_current}) yields the Alfven current $I_{\text{A}}$. 
These formulas have been obtained in the context of Z-pinches 
\cite{HainesJPD78}. In hot laser plasma $I_{\text{s}}$ can become much 
larger than $I_{\text{A}}$ since hollow cylinder-like density channels 
form and a significant lateral pressure is present. As a consequence, 
larger total currents than the Alfven current are possible in a single 
filament. From the simulation it is obtained that the singular current 
$I_{\text{s}}$ in the central filament of plot (b) in Fig. \ref{fig:jtevol} 
is about $100 \, \text{kA}$. Roughly this value is also obtained for
$I=\int dxdy \, j_{\text{z}}$ from the simulation by integrating over the 
cross sectional area of this filament. 

Figure \ref{fig:spectrum} shows the electron energy spectrum (a),
the integrated longitudinal and lateral charge and energy flows (b),
and the electron number and energy densities in the plasma slab (c). 
The longitudinal flows have been split into positive and negative 
components. The return energy flow is negligible. For the lateral 
flows the integrated absolute values are taken. The fastest electrons 
obtained in the simulation have about $30 \, \text{MeV}$ while fractional 
absorption of the laser is about $35 \, \%$. The absorbed power of the 
laser is transported away by the energy flow $q_{\text{z}}$ shown in 
plot (b). The latter consists predominantly of fast electrons. The 
energy flow penetrates much deeper into the plasma than the charge 
flow but still decays rapidly. Plot (c) shows that the energy density
is large only in the front layer of the plasma slab. Since the longitudinal 
and lateral charge flows are only sizeable where the magnetic field is 
large (see Fig. \ref{fig:bt}) the implication is that the mutual 
interaction between current and magnetic field confines most of the 
total charge and energy flows to the front surface of the slab. Since 
time-averaged electric fields in the simulation are much smaller than 
time-averaged magnetic fields the transport regime is of 
magneto-hydrodynamic nature.

In conclusion it has been shown that large charge flows generated by 
intense laser radiation decay into current filaments. The current 
filaments are isolated by magnetic filaments which reconnect until 
larger ones are obtained. This process proceeds until a dominating
filament is obtained in the plasma. The magnetic pressure associated
with the magnetic filaments leads to hollow density channels. Hollow
density channels, however, can support total currents that exceed the
Alfven current. This regime of transport is different from the one 
investigated by Bell {\em et al.} \cite{BellPPCF97}. Despite hollow 
channel transport severe charge and energy flow inhibition are observed. 
We note that the Alfven current limit is difficult to apply in the
context of a hot laser plasma since it has many ambiguities. We note 
that the plasma volume investigated is small due to numerical expense 
and that lateral boundaries are periodic. Hence, some care has to be 
taken when extrapolating the results reported here to FI relevant 
problems.

Many helpful discussions with M. G. Haines, F. Pegoraro, and P. Mulser
are acknowledged. The present work has been sup\-ported by the DFG
Schwerpunkt WECHSELWIRKUNG INTENSIVER LASERFELDER MIT MATERIE.
Use of the Supercomputing facilities at ZIB (Konrad Zuse, Berlin,
Germany) and NIC (John von Neumann Institute, J\"ulich, Germany) 
has been made.

\newpage

\begin{figure}
\epsfxsize=10cm
\caption[]{Merging of magnetic filaments and cycle averaged $E$ and $B$ 
fields vs $z$ in units $E_{\text{0}}$ and $B_{\text{0}}$. Plots (a,b,c) 
show the planes $z=2.0 \, \mu \text{m}$ of the cycle averaged magnetic 
field $B=(B^2_{\text{x}}+B^2_{\text{y}}+B^2_{\text{z}})^{0.5}$ at different 
times. Plots (d,e,f) show $E=\int dxdy \, (E^2_{\text{x}}+E^2_{\text{y}}+
E^2_{\text{z}})^{0.5}$ (dashed) and $B=\int dxdy \, (B^2_{\text{x}}+
B^2_{\text{y}}+B^2_{\text{z}})^{0.5}$ (solid) averaged over the lateral 
area and a laser cycle. The times are $t=71 \, \text{fs}$ (a,d), $t=125 \, 
\text{fs}$ (b,e), and $t=179 \, \text{fs}$ (c,f). The arrows in (a,b,c) 
indicate the direction of the cycle averaged magnetic field ${\bf B}$. 
The parameters are $E_{\text{0}}=1.0 \cdot 10^{11} \, \text{V/m}$ and 
$B_{\text{0}}=8.76 \cdot 10^{2} \, \text{Vs/m}^{2}$.}
\label{fig:bt}
\end{figure}

\begin{figure}
\epsfxsize=10cm
\caption[]{Cycle averaged $j_{\text{z}}$ in units $j_{\text{0}}$. 
Plots (a,b,c) show the planes $z=2.0 \, \mu \text{m}$ of the cycle 
averaged current density $j_{\text{z}}$ at different times. Plots 
(d,e,f) show $I=\int dxdy \, j_{\text{z}}$ split into flow 
(dashed-dotted) and return flow (solid) averaged over a laser 
cycle. The total current in the central filament (blue color) 
is approximately $200 \, \text{kA}$. The times are $t=71 \, \text{fs}$ 
(a,d), $t=125 \, \text{fs}$ (b,e), and $t=179 \, \text{fs}$ (c,f). 
The arrows in (a,b,c) indicate the direction of the cycle averaged 
magnetic field ${\bf B}$. The parameter is $j_{\text{0}}=1.66 \cdot 
10^{16} \, \text{A/m}^2$.}
\label{fig:jtevol}
\end{figure}

\begin{figure}
\epsfxsize=10cm
\caption[]{Hollow density channels. Cycle averaged electron density 
$n_{\text{e}}$ in units $n_{\text{0}}$. Plots (a,b,c) show the planes 
$z=2.0 \, \mu \text{m}$ at different times. Plot (d) shows the plane 
$x=2.0 \, \mu \text{m}$. The times are $t=71 \, \text{fs}$ (a), $t=125 
\, \text{fs}$ (b), and $t=179 \, \text{fs}$ (c,d). The parameter is 
$n_{\text{0}}=1.0 \cdot 10^{28} \, \text{m}^{-3}$. The location of 
the hollow density channels coincides with magnetic filaments as is
seen by comparing with plots (a,b,c) of Fig. \ref{fig:bt}.}
\label{fig:ne}
\end{figure}

\begin{figure}
\epsfxsize=7cm
\caption[]{Energy spectrum (a), total current and energy flows in units 
$I_{\text{0}}$ and $q_{\text{0}}$ (b), total electron and electron
energy density in units $n_{\text{0}}$ and $\epsilon_{\text{0}}$ (c). Plot 
(a) shows the electron energy spectrum obtained from all the particles 
in the plasma. The fastest electrons have $30 \, \text{MeV}$. The thin 
solid lines of plot (b) show the integrated current and return current 
$I=\int dxdy \, j_{\text{z}}$. The thick solid line indicates $I=\int 
dxdy \, \sqrt{j^2_{\text{x}}+j^2_{\text{y}}}$. The thin dashed-dotted 
lines indicate the energy and return energy flows $q_z=\int dxdy \, 
\int d^3p \, v_z \, (c\sqrt{m^2c^2+{\bf p}^2}-mc^2) \, f$. The thick 
dashed-dotted line shows the lateral energy flow $q_{\perp}=\int dxdy \, 
\int d^3p \, \sqrt{v^2_{\text{x}}+v^2_{\text{y}}} \, (c \sqrt{m^2c^2+
{\bf p}^2}-mc^2) \, f$. The solid line of plot (c) shows the integrated 
electron energy density $\epsilon=\int dxdy \, \int d^3p \, (c\sqrt{m^2c^2
+{\bf p}^2}-mc^2) \, f$. The dashed-dotted line gives the electron number 
density $n_{\text{e}} \rightarrow \int dxdy \, n_{\text{e}}$. Fractional 
absorption is about $35.0 \, \%$. The time is $t=179 \, \mbox{fs}$.
The parameters are $I_{\text{0}}=10^5 \, \mbox{A}$, $q_{\text{0}}=10^{12} 
\, \mbox{W}$, $\epsilon_{\text{0}}=10^4 \, \mbox{J/m}$, and $n_{\text{0}}
=10^{17} \, \mbox{m}^{-1}$.}
\label{fig:spectrum}
\end{figure}

\pagestyle{empty}
\unitlength1cm





\end{document}